\documentclass[%
 reprint,
 amsmath,amssymb,
 aps,
]{revtex4-1}

\usepackage{graphicx}
\usepackage{dcolumn}
\usepackage{bm}

\begin{document}

\preprint{APS/123-QED}

\title{Electric Conductivity from the solution of the Relativistic Boltzmann Equation}

\author{A. Puglisi}
\email{puglisia@lns.infn.it}
\author{S. Plumari}%
\email{salvatore.plumari@ct.infn.it}



\author{V. Greco}
\email{greco@lns.infn.it}
\affiliation{Department of Physics and Astronomy, University of Catania, Via S. Sofia 64, I-95125 Catania, Italy}
\affiliation{Laboratorio Nazionale del Sud, INFN-LNS, Via S. Sofia 63, I-95125 Catania, Italy }


\date{\today}

\begin{abstract}
We present numerical results of electric conductivity $\sigma_{el}$ of a fluid obtained solving the Relativistic Transport Boltzmann equation in a box with periodic boundary conditions.
We compute $\sigma_{el}$ using two methods: the definition itself, i.e. applying an external electric field, 
and the evaluation of the Green-Kubo relation based on the time evolution of the current-current correlator. 
We find a very good agreement between the two methods.

We also compare numerical results with analytic formulas in Relaxation Time Approximation (RTA) where the
relaxation time for $\sigma_{el}$ is determined by the transport cross section $\sigma_{tr}$, i.e. the
differential cross section weighted with the collisional momentum transfer. We investigate the electric conductivity dependence on the microscopic details of the 2-body scatterings: isotropic and anisotropic cross-section, and massless and massive particles. We find that the RTA underestimates considerably $\sigma_{el}$; for example at
screening masses $m_D \sim \,T$ such underestimation can be as large as a factor of 2.
Furthermore, we study a more realistic case for a quark-gluon system (QGP) considering both a quasi-particle model,
tuned to lQCD thermodynamics, as well as the case of a pQCD gas with running coupling. 
Also for these cases more directly related to the description of the QGP system, we find that RTA 
significantly underestimate the $\sigma_{el}$ by about a $60-80\%$.

\end{abstract}

\pacs{Valid PACS appear here}

\maketitle

\section{Introduction}
Relativistic Heavy Ions Collision experiments performed by Relativistic Heavy Ion Collider (RHIC) at BNL and  Large Hadron Collider (LHC) at CERN have produced a Little Bang, meaning that the same conditions of temperature and energy density of early universe have been generated.
A system of strongly interacting particles above the critical temperature $T_c\sim 160\,MeV$ \cite{Science_Muller, Shuryak:2003xe} is expected to undergo to a phase transition from hadron matter to Quark-Gluon Plasma (QGP) \cite{lQCD, Bazavov:2011nk}.
The collective behaviour observed in experiments \cite{Adams} and theoretical and phenomenological studies of viscous hydrodynamics \cite{Romatschke:2007mq,Song:2011hk,Schenke:2010nt,Niemi:2011ix} and parton transport \cite{Ferini:2008he,Plumari_Bari,Xu:2007jv,Xu:2008av,Cassing_JCP,Cassing:2009vt,Bratkovskaya:2011wp} have confirmed that QGP behaves like a fluid with a very small shear viscosity to entropy density ratio $\eta/s$ close to the lower bound $1/4\pi$ predicted by AdS/CFT \cite{Kovtun:2004de}.
This suggests that QGP could be a nearly perfect fluid with the smallest viscous dynamics ever observed, even less dissipative than the ultra cold matter created by magnetic traps \cite{Cao:2010wa,O'Hara:2002zz}.
Being the QGP created at HICs a system far from equilibrium, the study of its transport coefficients is attracting a  great interest. The $\eta$ has been studied extensively \cite{Kapusta_qp,Sasaki,Khvorostukhin,Sasaki_njl,Wiranata, Fuini:2010xz,Muronga,Wesp:2011yy,Plumari_visco,Cassing_GK}. Only very recently the electric conductivity,  that represents the response of a system to the applied electric field, has captured a significative importance in the  field of strongly interacting matter for many motivations. Electric conductivity $\sigma_{el}$ can be computed on the Lattice from correlation function. On the experimental side, Heavy Ion Collisions are able to produce very strong electric and magnetic fields  ($eE\simeq eB\simeq m_{\pi}^2$, with $m_{\pi}$ the pion mass) in the first $1-2\,fm/c$ from the collision \cite{Tuchin,Hirono,Gursoy:2014aka,McLerran:2013hla}. The value of $\sigma_{el}$ would be of fundamental importance for the strength of the Chiral-Magnetic Effect \cite{Fukushima}, a signature of the CP violation of the strong interaction. It has also  been shown that in mass asymmetric collision the electric field has a privileged direction generating a current whose effects can be observed in collective flow and are directly related to $\sigma_{el}$ \cite{Hirono}.
Moreover $\sigma_{el}$ can be related to the emission rate of soft photons \cite{Kapusta_book} accounting for their raising spectra \cite{Turbide:2003si,Linnyk:2013wma}.
In this work we compute electric conductivity solving numerically the Relativistic Boltzmann Transport equation with a transport code already developed to study the dynamics of heavy-ion collisions at RHIC and LHC energies\cite{Ferini:2008he, Greco, Plumari_njl, Plumari_Bari, Ruggieri, Scardina} considering two body elastic cross-section.
Electric conductivity can be calculated using two methods. The first represents the empirical method of measuring the electric conductivity and is suggested just by the operative definition of this transport coefficient $\vec{j}=\sigma_{el} \vec{E}$: applying an external electric field to a system and evaluating the electric current induced one can obtain the proportionality coefficient $\sigma_{el}$.
The other method comes from the Linear Response Theory and consists in employing the Green-Kubo correlator: in this case electric conductivity is related to the time-correlation function of the current evaluated in thermal equilibrium, i.e. without any external electric field applied.
We found that the two methods are in very good agreement in the temperature range explored.
We study the microscopic dependence of $\sigma_{el}$ on the particular scattering, isotropic or anisotropic cross-section, exploring also massless and massive particles showing a very similar behavior of shear viscosity $\eta$, as already found in our previous work \cite{Plumari_visco} for massless case.
We compare numerical results with analytical formulas obtained in Relaxation Time Approximation (RTA) and  we find that RTA is a quite good approximation for the simplest case of massless particles interacting via isotropic cross-section even if it overestimates numerical results of about $30\%$ for temperature $T>0.4\,GeV$.
For anisotropic cross-section and massive particles RTA underestimates more considerably electric conductivity than shear viscosity.

Being interested in a more realistic case and in a comparison with Lattice QCD results, we employed a quasi-particle model to take into account the thermodynamics of strongly interacting matter \cite{Plumari_qpmodel}. We consider a system of massive quasi-quarks and quasi-gluons interacting with a pQCD-like cross-section temperature and energy dependent. We also calculate electric conductivity in the limit of massless particles with a running coupling constant from pQCD calculation to have an asymptotic value of $\sigma_{el}$ for $T\gg T_c$ and to study the dependence on a different coupling.

The paper is organized as follows. In Sec. \ref{Sec:Electric_conductivity} we briefly recall Transport Theory and Relaxation Time Approximation to obtain the analytical formula of electric conductivity $\sigma_{el}$ and also Green-Kubo relation. In Sec. \ref{Sec:box_setup} we present the setup of our numerical simulations solving numerically the Relativistic Transport Boltzmann equation and how to compute $\sigma_{el}$ in this framework.
In Sec. \ref{Sec:el_cond_iso} we show numerical results of $\sigma_{el}$ for the most simple system: massless particles interacting with isotropic and constant cross-section.
In Sec. \ref{Sec:el_cond_anisotropic} we consider a more general case of massive particles and anisotropic scatterings for both $\sigma_{el}$ and $\eta$.
In Sec. \ref{Sec:el_cond_quasi_particle} we discuss $\sigma_{el}$ results for a  more realistic system of a quasi-particle model, that describes the thermodynamics of Lattice QCD. Furthermore we show the pQCD results.
 
\section{\label{Sec:Electric_conductivity}Electric Conductivity}

The electric conductivity $\sigma_{el}$ represents the response of a system to an external electric field.
The definition of $\sigma_{el}$ comes from the microscopic version of Ohm's Law:
\begin{equation}
\vec{j}=\sigma_{el} \vec{E}.
\end{equation}
The starting point of our calculation is the Relativistic  Boltzmann Transport (RBT) equation that in the presence of an external field can be written as \cite{Yagi, Cercignani}:
\begin{equation} \label{Boltzmann_eq}
p^{\mu}\partial_{\mu} f(x,p) + q F^{\alpha\beta}p_{\beta} \frac{\partial}{\partial p^{\alpha}} f(x,p) = {\cal C}[f] 
\end{equation}
where $f(x,p)$ is the distribution function, $F^{\alpha\beta}$ is the electromagnetic field strength tensor, $C[f]$ is the collision integral which, considering only $2\to 2$ scatterings, can be written as follows
\begin{equation}
{\cal C}(x,p)=\int_{2}\int_{1'}\int_{2'} \left( f_{1'}f_{2'} - f_1 f_2 \right) |{\cal M}_{1' 2' \to 12}|\delta^{4}(p_1+p_2-p_1' p_2')
\end{equation}
where ${\cal M}$ is the transition matrix for the elastic process linked to the differential cross-section $|{\cal M}|^2=16\pi s(s-4M^2)d\sigma/dt$. The possible extension to $2\leftrightarrow 3$ processes has been thoroughly discussed in Ref. \cite{Wesp:2011yy}.
In order to obtain an analytical solution for the Boltzmann equation, it is necessary to approximate the Collision integral. The most simple scheme is the Relaxation Time Approximation which assumes the following form:
\begin{equation}\label{C_RTA}
{\cal C}[f] \simeq -\frac{p^{\mu}u_{\mu}}{\tau} (f-f_{eq})
\end{equation}
where $u^{\mu}$ is the velocity flow that in local rest frame is $(1,\textbf{0})$, $\tau$ is the so-called relaxation time which determines the time scale for the system to relax toward the equilibrium state characterized by $f_{eq}$.
Assuming that the distribution function $f$ is near the equilibrium one $f_{eq}$, one can write:
\begin{equation}
f(x,p,t)=f_{eq}(x,p)(1+\phi).
\end{equation}
If we consider a uniform system and only an electric field $\vec{E}$, from Eq. (\ref{Boltzmann_eq}) and Eq. (\ref{C_RTA}) we obtain:
\begin{equation}
-q\left( p_0 \vec{E} \cdot \frac{\partial f_{eq}}{\partial \vec{p}} - \vec{E}\cdot \vec{p} \frac{\partial f_{eq}}{\partial p^0} \right) = -\frac{p^0}{\tau} f_{eq}\phi
\end{equation}
and solving for $\phi$, assuming $\phi \ll f_{eq}$, one obtains:
\begin{equation}
\phi=\frac{1}{T}q\tau\, \vec{E}\cdot \frac{\vec{p}}{p^0} 
\end{equation}
The electric current is:
\begin{equation}
j^{\mu}=q\int \frac{d^3p}{(2\pi)^3} \frac{p^{\mu}}{p^0}\,f=q\int \frac{d^3p}{(2\pi)^3}\frac{p^{\mu}}{p^0}\,f_{eq}(1+\phi)
\end{equation}
Using $\phi$ of the previous equation, considering the definition of electric conductivity and generalizing to a system of different charged particles one obtains \cite{FernandezFraile:2005ka, Puglisi}:
\begin{equation}\label{el_cond_formula}
\sigma_{el}=\frac{e^2}{3 T} \sum_{j=q,\bar{q}} q_j^{2}\int\frac{d^3p}{(2\pi)^3} \frac{\vec{p}^2}{E^2}  \tau_j f_{eq} = \frac{e_{\star}^2}{3 T} \left\langle \frac{\vec{p}^2}{E^2} \right\rangle \tau_{q}\rho_{q}
\end{equation}
where $q_j$ is the quarks charge ($\pm 1/3,\pm 2/3$), $\tau_j$ is the relaxation time for quarks, $\rho_q$ is the quark  density, $e_{\star}^{2}=e^2\sum_{j=\bar{q},q}q_j^2=4e^2/3$.
We notice that in the classical limit Eq. (\ref{el_cond_formula}) simplifies to the well known Drude formula $\tau e^2 \rho/m$ while in the ultrarelativistic limit becomes $\tau e^2/3T$.

Transport coefficients can be computed in a more general way in the framework of  Linear Response Theory where each coefficients can be related, according to Green-Kubo relations, to correlation functions of the corresponding flux or tensor in thermal equilibrium.
Green-Kubo formula for electric conductivity has the following form:
\begin{equation}\label{GK_el_cond}
\sigma_{el}=\frac{V}{T}\int_{0}^{\infty}dt\, \langle j_{z}(t)j_z(0) \rangle=\frac{V}{T} \langle j_z(0)j_z(0) \rangle \tau_{\sigma_{el}}
\end{equation}
where $j_z$ is the $z$ component of electric current, $\langle \dots \rangle$ is the thermal average at equilibrium. We have also assumed a uniform system so the volume integral gets a $V$ factor, the right hand side is obtained using the fact that the time-correlation function for a system in thermal equilibrium is a decreasing exponential $\exp(-t/\tau)$ \cite{landau}. Such a behaviour has been carefully checked to be satisfied by our numerical solution of RBT, similarly to Ref. \cite{Plumari_visco} for the shear viscosity.
We notice that the numerical evaluation of Green-Kubo correlator is a powerful method for computing, within the validity of Linear Response Theory,  transport coefficients without any kind of approximation being the only requirement the thermal equilibrium. In Ref. \cite{Plumari_visco} we have showed the numerical convergency of Green-Kubo method in the framework of transport code.

\section{\label{Sec:box_setup} Box setup}

We solve numerically the transport equation using the stochastic interpretation of the transition amplitude \cite{Xu:2008av,Cassing_JCP}, according to which collision probability of two particles in a cell of volume $\Delta V_{cell}$ and a time-step $\Delta t$ is
\begin{equation}
P=v_{rel}\frac{\sigma_{tot}}{N_{test}}\frac{\Delta t}{\Delta V_{cel}}
\end{equation}
where $\sigma_{tot}$ is the total cross section and $v_{rel}=\sqrt{s(s-4M^2)}/2E_1 E_2$ is the relative velocity of the two incoming particles.

The definition of electric conductivity $J=\sigma_{el}E$ suggests the experimental method for evaluating the electric conductivity simply inverting the relation $\sigma_{el}=J/E$: taking the ratio between the electric current measured and the electric field applied one obtains $\sigma_{el}$. Of course one needs also to verify that $J/E$ is independent on $E$ itself. In the following discussion we will call this method the \textit{E-field method}. We notice that recently such a method has been employed to evaluate $\sigma_{el}$
in the parton-hadron-string dynamics (PHSD) transport approach \cite{Cassing_el}

To simulate a constant electric field $\vec{E}$ in the box, it is sufficient to modify the equation of motion of each particle as follows:
\begin{equation}
\frac{d}{dt}p^i_z = q_i e E_z
\end{equation}
where $q_j$ is the charge of the particle and we have chosen the electric field along the $z-$direction.
The electric current in the $z$ direction for a discrete system of particles has the following form:
 \begin{equation}
 j_z(t)=\frac{1}{V} \sum_i \frac{e q_i p^i_z(t)}{p_0^i}
 \end{equation}
 where the sum is over  particles, $p_0$ is the particle energy and $V$ is the volume of the system.

We performed simulations in a uniform box  of volume $V=5^3\,fm^3$ with periodic boundary conditions using time step $\Delta t=0.01\,fm/c$, spatial discretization $\Delta V_{cell}=0.1^3\,fm^3$ and $N_{test}\times N_{real}\sim 500\,10^3$  that ensures the numerical convergency; we have checked that results are independent on volume size. 
Particles are distributed uniformly in space and according to Boltzmann distribution function, $f(p)=e^{-E/T}$, in momentum.

\begin{figure}
 \centering
 \includegraphics[scale=0.3, keepaspectratio=true]{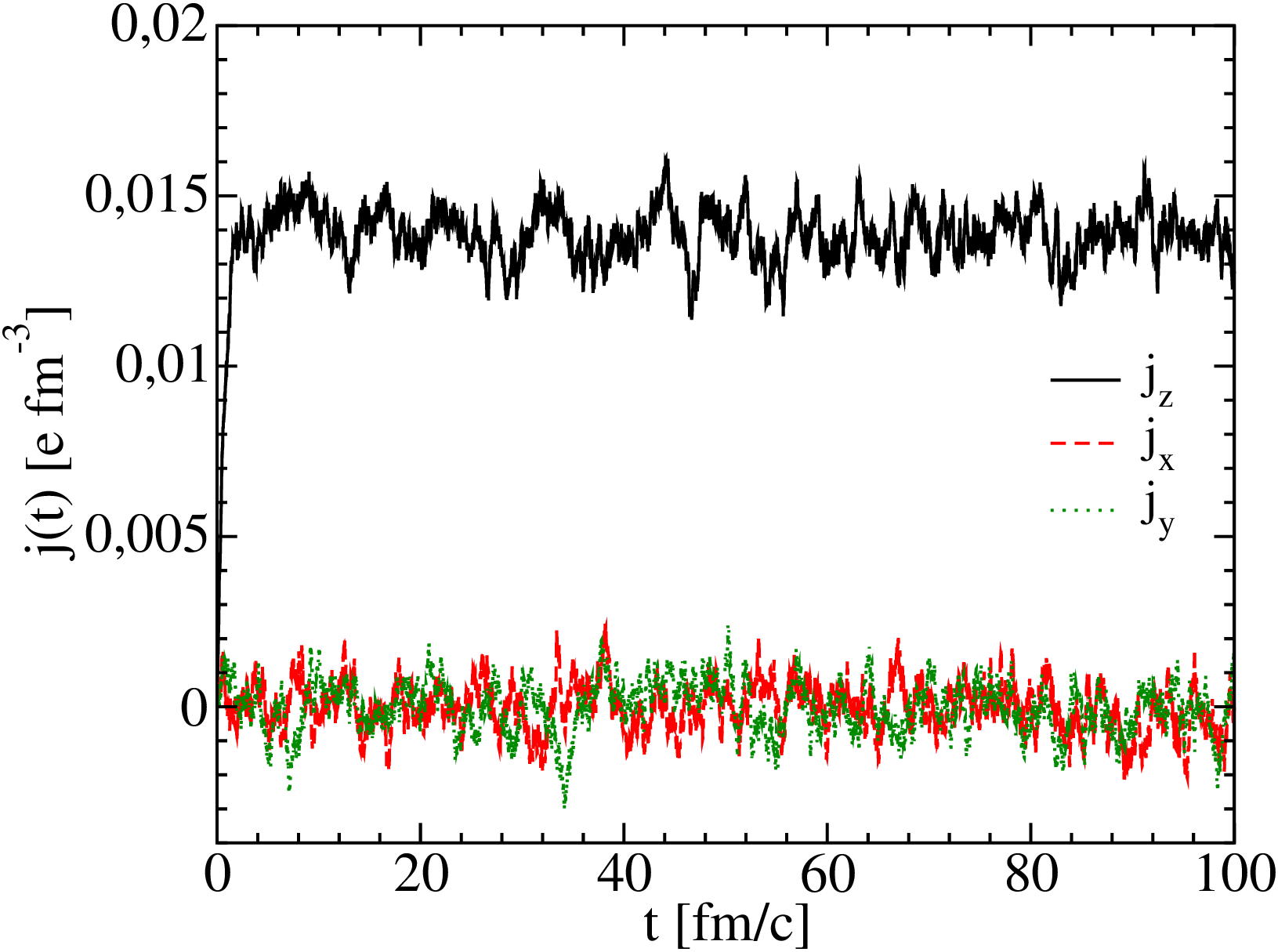}
 \caption{$x,y,z$ components of electric current $j$ as a function of time for  electric field $eE=0.05\,GeV/fm$ in the $z$ direction. We fixed  $T=0.2\,GeV$, $m=0.4\,GeV$ and $\sigma_{tot}=10\,mb$ for all particles in this simulation. $j_z$ reaches a saturation value proportional to $E$ while $x$ and $y$ components fluctuate around $0$.}
 \label{fig:jz_E}
\end{figure}

In Fig. \ref{fig:jz_E} we show an example of $x,y,z$ components of the electric current $j(t)$ as a function of time. In this simulation we consider a system of massive quarks, antiquarks and gluons ($m=0.4\,GeV$) at thermal and chemical equilibrium interacting with isotropic cross-section $\sigma_{tot}=10\,mb$ and we have applied an electric field $eE=0.05\,GeV/fm$ in the $z$ direction. 
As we can see the $z$ component (black solid line) reaches a saturation value while $x$ and $y$ components fluctuate around the equilibrium value zero.

\begin{figure}
 \centering
 \includegraphics[scale=0.3, keepaspectratio=true]{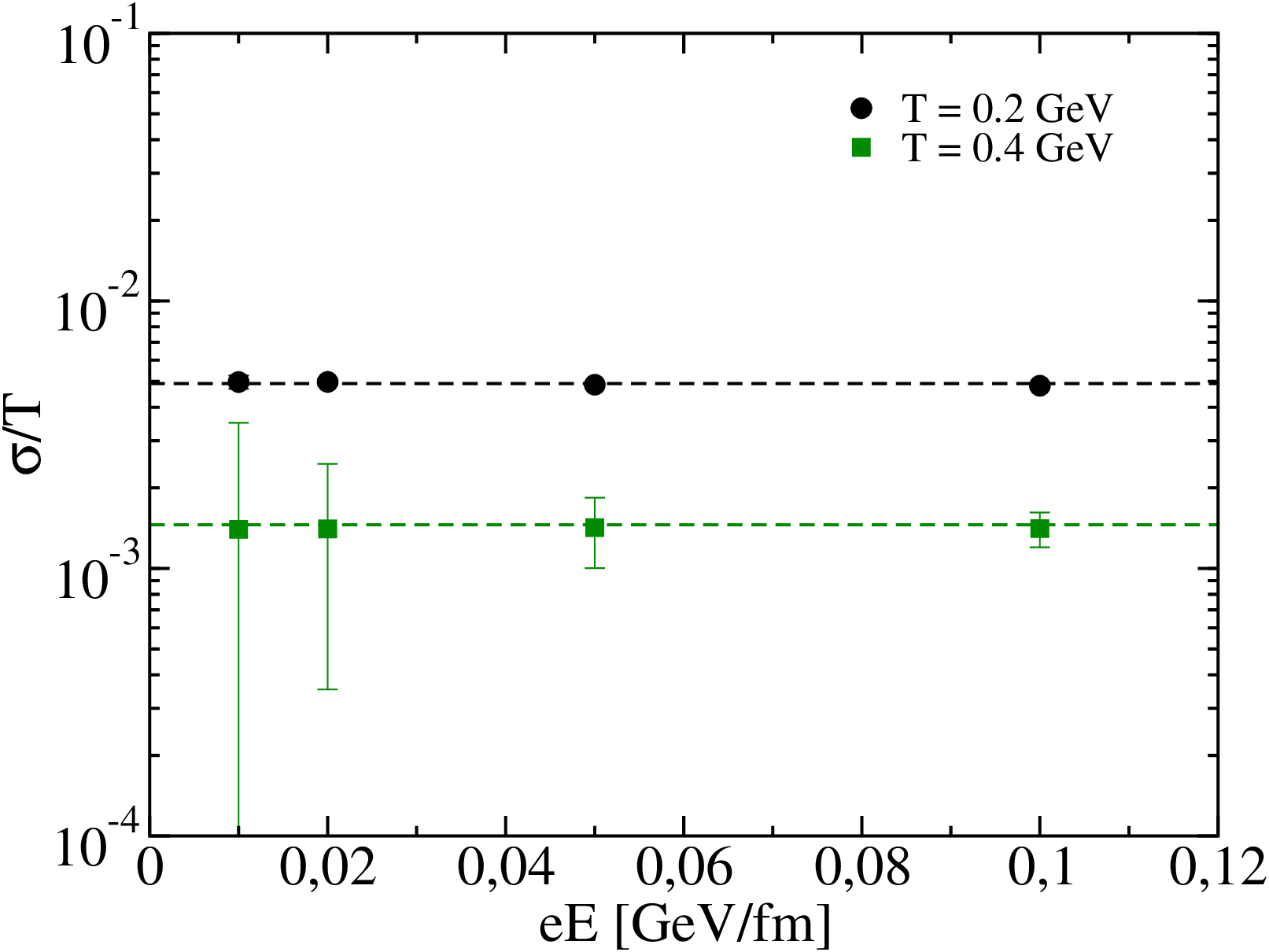}
 \caption{Electric conductivity $\sigma_{el}/T$ as a function of electric field. We fixed  $T=0.2\,GeV$ (dark circles) and $T=0.4\,GeV$ (green squares), $m=0.4\,GeV$ and $\sigma_{tot}=10\,mb$ isotropic for all particles in these simulations. Results are compatible with a constant ratio $\sigma_{el}/T$ in the range of electric field explored. Dashed lines are RTA predictions.}
 \label{fig:sigma_E}
\end{figure}

In Fig. \ref{fig:sigma_E} is shown the ratio $\sigma_{el}/T$ as a function of the applied electric field $eE$ for two different temperatures $T=0.2 \,GeV$ dark circles and $T=0.4\,GeV$ green squares for a system of massive particles with $m=0.4\,GeV$ interacting via isotropic cross-section with $\sigma_{tot}=10\,mb$: $\sigma_{el}/T$ is independent on the applied electric field which confirms the validity of its definition. Dashed lines are RTA predictions of Eq. (\ref{el_cond_formula}). As we can see from Fig. \ref{fig:sigma_E} thermal fluctuations affect the uncertainties on electric conductivity because a greater temperature produces greater fluctuations in the saturation value of electric current.  The increasing of the electric field has the effect of developing a more stable electric current that is easily noticeable in the decreasing of error bars. However, the electric field cannot be increased arbitrarily because one has to guarantee the linear response of the system: with a very high value of $eE$ the system could not reach any equilibrium value of electric current and the definition itself of $\sigma_{el}$ becomes meaningless.
We have checked the correct behaviour of electric current for several value of electric field, temperature and cross-section presented in this work.

\begin{figure}
 \centering
 \includegraphics[scale=0.3, keepaspectratio=true]{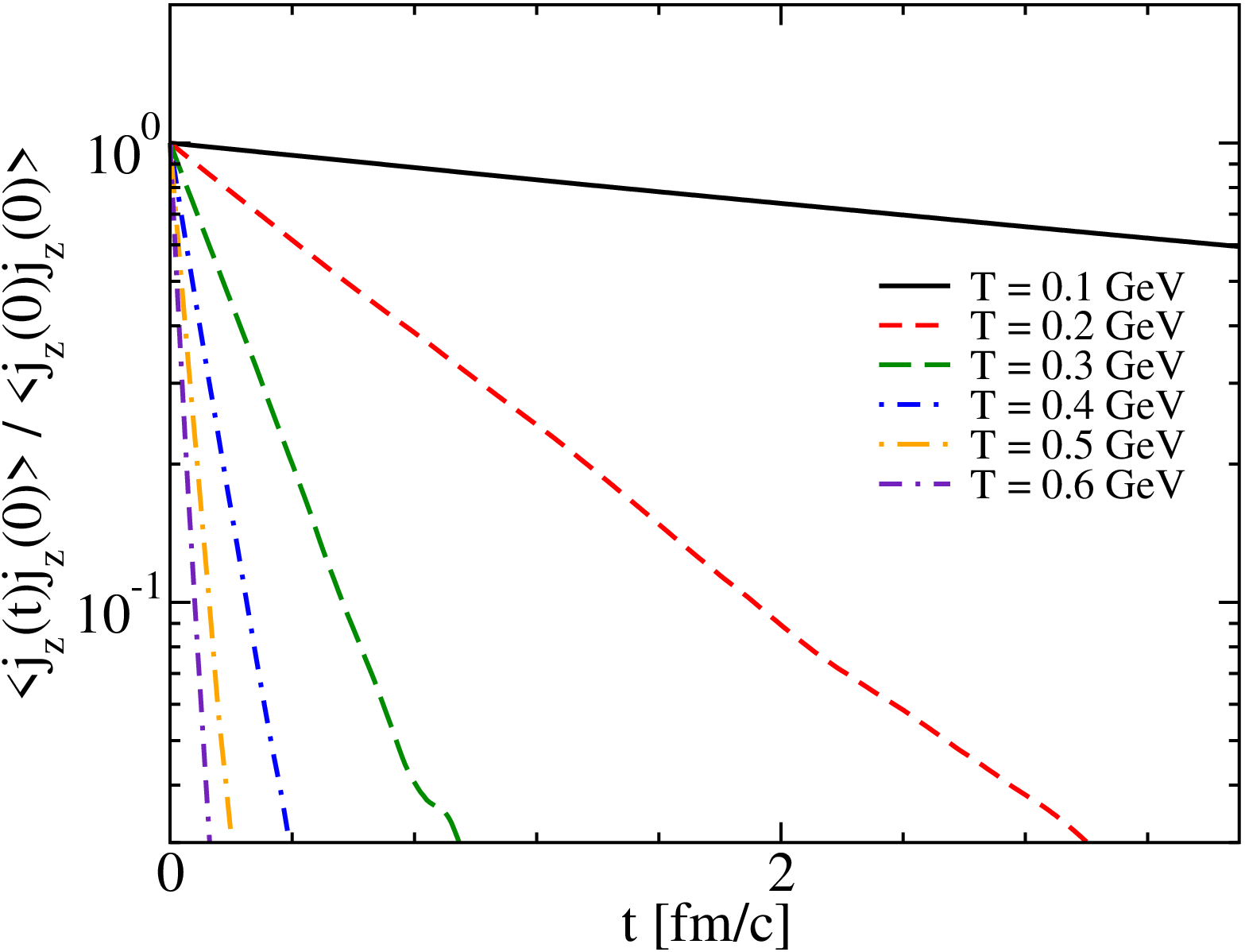}
 \caption{Green-Kubo correlators $\langle j_z(t) j_z(0)\rangle$ normalized to the initial value $\langle j_z(0) j_z(0) \rangle$ for a system of massless quarks and gluons interacting with isotropic cross-section ($\sigma_{tot}=3\,mb$) for temperature $T=0.1-0.6\,GeV$. We can see the typical behaviour of a decreasing exponential $\exp(-t/\tau)$.}
 \label{fig:GK-correlators}
\end{figure}

The other method we used to compute $\sigma_{el}$ is the evaluation of Green-Kubo formula Eq (\ref{GK_el_cond}). The correlation function $\langle j_z(t)j(0) \rangle$ can be written as follows:
\begin{align}
\langle j_z(t)j(0) \rangle &= \left\langle  \lim_{T_{max}\to \infty} \frac{1}{T_{max}} \int_{0}^{T_{max}} dt'\,j_{z}(t+t')j(t')  \right\rangle =\\
&= \left\langle \frac{1}{N_{T_{max}}} \sum_{j=1}^{N_{T_{max}}} j_z(i\Delta t+ j \Delta t) j_z(j\Delta t) \right\rangle
\end{align}
where $T_{max}$ is the maximum time chosen in our simulations, $N_{T_{max}}=T_{max}/\Delta t$ represents the maximum number of time-steps and $i\Delta t=t$, while $\langle \cdots \rangle$ denotes the average over events generated numerically.
In Fig. \ref{fig:GK-correlators} we plot an example of Green-Kubo correlation functions normalized to the initial value $\langle j_z(0)j_z(0) \rangle$ for a system of massless quarks and gluons interacting via isotropic cross-section ($\sigma_{tot}=3\,mb$) for several temperatures in the range $T=0.1-0.6 \,GeV$. Correlation functions $\langle j_z(t)j_z(0) \rangle$ behave like decreasing exponential $\exp(-t/\tau)$ as it should be for a system in thermal equilibrium. As one increases the temperature, the slope $\tau$ decreases as expected from kinetic theory $\tau\sim 1/(\rho(T) \sigma)$.

We recall that for the calculation of $\sigma_{el}$ and $\eta$ using Green-Kubo relation the setup of the box is simply in thermal equilibrium, i.e. without any external electric field.

\section{\label{Sec:el_cond_iso} Electric conductivity: isotropic scattering}
In this section we consider the simplest case of a system of massless particles (quarks, anti-quarks and gluons) interacting via isotropic and elastic scatterings.
In this case the transport relaxation time has the following form:
\begin{equation}
\tau_{tr,i}^{-1}=\sum_{j=q,\bar{q},g} \langle \rho_j v_{rel}^{ij} \sigma_{tr}^{ij} \rangle=
\frac{2}{3}\sigma_{tot} (\rho_q + \rho_{\bar{q}} + \rho_g)
\end{equation}
where $v_{rel}$ is the relative velocity of the two incoming particles and  for massless particles $v_{rel}=1$, $\sigma_{tr}$ is the transport cross-section that for isotropic scatterings is equal to $\frac{2}{3}\sigma_{tot}$, $\rho_{q,\bar{q},g}$ is respectively quarks, antiquarks and gluons density. Assuming that all particles interact with the same cross-section, Eq. (\ref{el_cond_formula}) simplifies as follows:
\begin{equation}\label{el_iso}
\frac{\sigma_{el}}{T}=\frac{e_{\star}^2}{3T^2} \frac{\gamma_q}{6\gamma_q+\gamma_g}\frac{1}{\frac{2}{3}\sigma_{tot}}.
\end{equation}
where $6$ comes from the sum over quarks flavour ($u,d,s,\bar{u},\bar{d},\bar{s}$) and we used $ \rho_{q,g}=\gamma_{q,g} T^3/\pi^2$, being  $\gamma$ the degeneracy factor.

\begin{figure}
 \centering
 \includegraphics[scale=0.3, keepaspectratio=true]{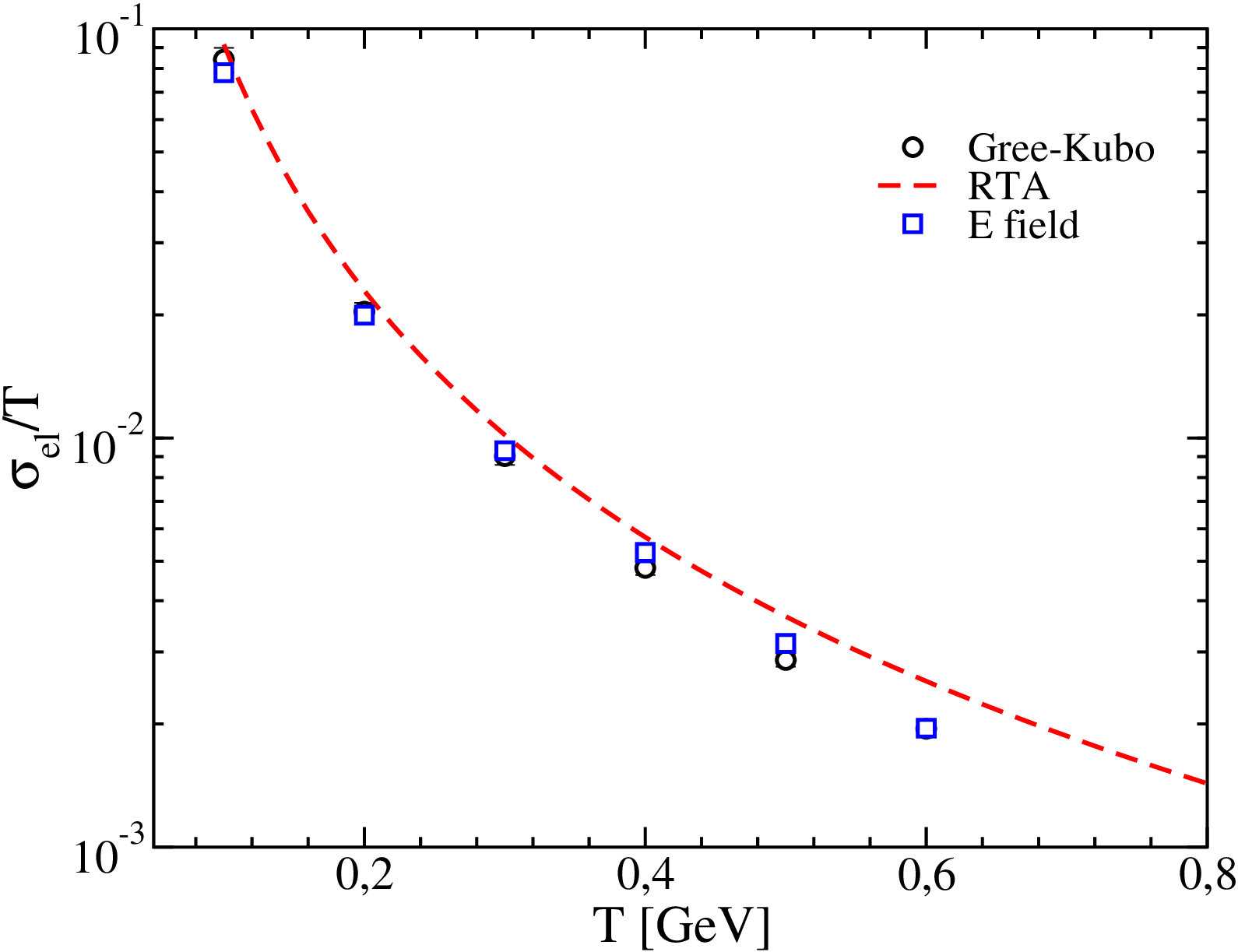}
 \caption{Electric conductivity $\sigma_{el}/T$ as a function of Temperature $T$ for a system of massless particles interacting via isotropic scatterings ($\sigma_{tot}=3\,mb$). Circles are Green-Kubo results while  blue squares represent the E-field method results. Red dashed line is Relaxation Time Approximation Eq. (\ref{el_iso}).}
 \label{fig:el_cond_iso}
\end{figure}

In Fig. \ref{fig:el_cond_iso} we show electric conductivity $\sigma_{el}/T$ as a function of temperature for a system of massless quarks, antiquarks and gluons interacting via the same isotropic cross-section $\sigma_{tot}=3\,mb$: open circles are computed using Green-Kubo relation Eq. (\ref{GK_el_cond}) while blue open squares are obtained with the E-field method ($eE=0.01-1.0\, GeV/fm$ for $T=0.1-0.6\,GeV$). Red dashed line represents Eq (\ref{el_iso}).
We can see a very good agreement between Green-Kubo and E-field method as it should be in the framework of Linear Response Theory: both definition of $\sigma_{el}$ are meaningful according to our results.
However RTA tends to overestimate numerical results in particular for high temperatures, e.g. at $T=0.6\,GeV$ there is a $30\%$ of discrepancy. This shows that the relaxation time $\tau_{\sigma_{el}}$ for the electric conductivity is only approximatively determined by the transport cross-section $\sigma_{tr}$. We will show in the following that the discrepancy between $\tau_{tr}$ and $\tau_{\sigma_{el}}$ becomes more drastic in the more general case of non isotropic scatterings.

\section{\label{Sec:el_cond_anisotropic}Electric conductivity and shear viscosity: anisotropic scattering}

In this section we consider  a quark-gluon plasma of massive quarks, antiquarks and gluons interacting via anisotropic cross-section.
In particular, we use the elastic pQCD inspired cross section with the infrared singularity regularized by 
Debye thermal mass $m_D$ \cite{Zhang}:
\begin{equation}\label{sigma_anis}
 \frac{d\sigma}{dt} = \frac{9\pi \alpha_s^2}{2}\frac{1}{\left(t-m_D^2\right) ^2}\left(1+\frac{m^2_D}{s}\right) 
\end{equation}
where $s,t$ are the Mandelstam variables.
This kind of cross-section is typically used in transport codes \cite{Xu:2008av,Ferini:2008he,Greco,Lin:2004en,Zhang:1999,Molnar:2002}.
In our calculations the Debye mass $m_D$ and the strong coupling constant 
$\alpha_{s}$ are both constant parameters. The total cross-section $\sigma_{tot}=9\pi \alpha_s^2/(m_{D}^2)$ with 
the above prescription is energy and temperature independent. 
The Debye mass is a parameter that establishes the anisotropy of the collision, for small values of $m_D$ we have that the 
distribution of the collision angle has a peak at small angles while in the opposite limit $m_D \gg T$, we recover the 
isotropic case.
For a fixed total cross-section, the transport cross-section $\sigma_{tr}$ can be written as
\begin{equation}
\sigma_{tr} (s)=\int \, \frac{d\sigma}{dt} \, \sin^{2}\Theta \, dt = \sigma_{tot} \, h(a) 
\end{equation}
where $h(a)=4 a ( 1 + a ) \big[ (2 a + 1) ln(1 + 1/a) - 2 \big ]$ and $a=m_{D}^2/s$. For $m_{D}\to \infty$ the function $h(a) \to 2/3$ 
and $\sigma_{tr}=(2/3)\sigma_{tot}$ we recover the isotropic limit, see the previous Sec.,  while for finite value of $m_D$ the function $h(a)<2/3$. 

\begin{figure}
 \centering
 \includegraphics[scale=0.3, keepaspectratio=true]{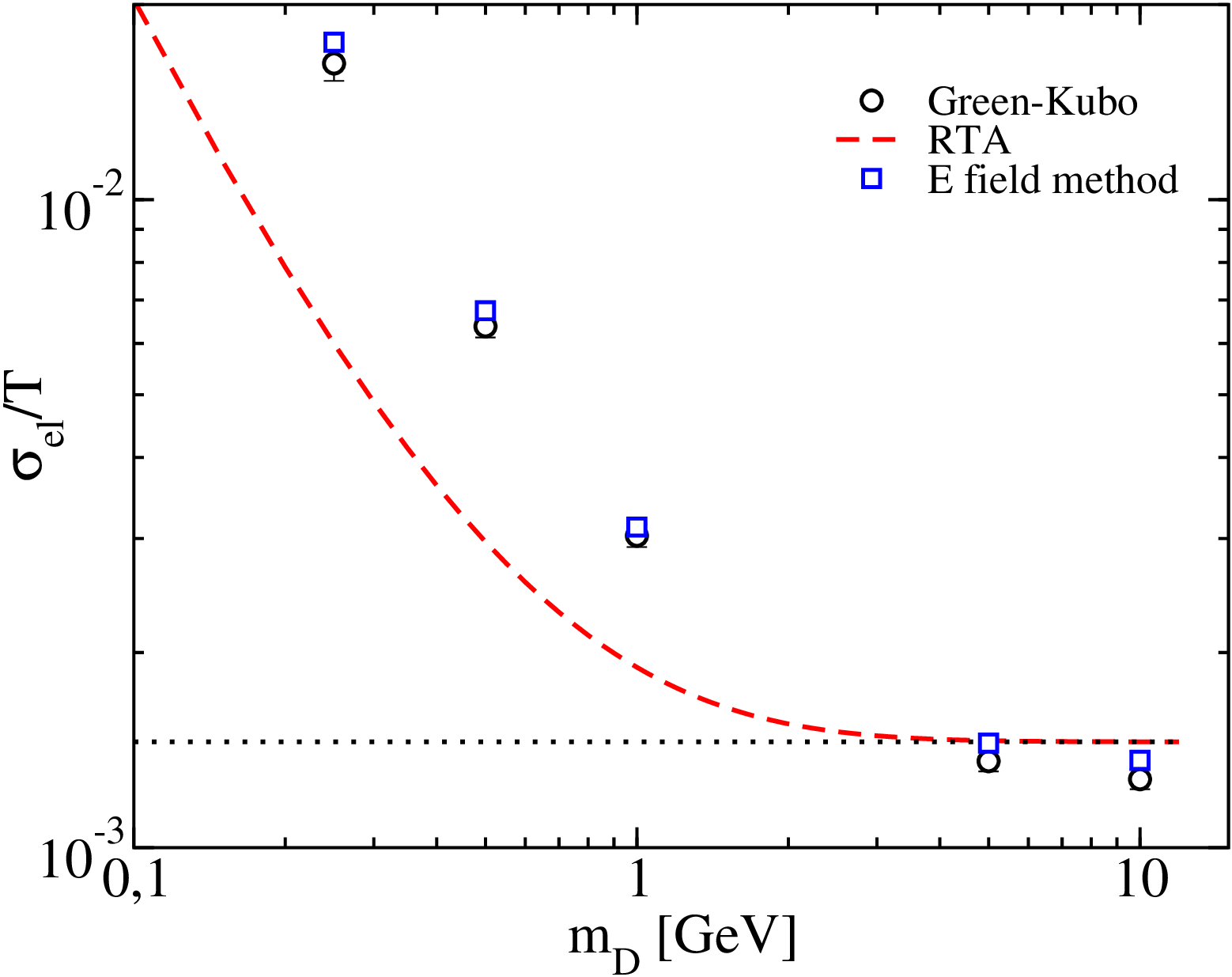}
 \caption{Electric conductivity as a function of Debye mass $m_D$.
In this simulations we set  $T=0.4\,GeV$, $\sigma_{tot}=10\,mb$ and $m=0.4\,GeV$ for all particles. Circles are Green-Kubo results. 
Red dashed line is RTA Eq. (\ref{el_cond_formula}) with $\tau_{tr}$ from Eq. (\ref{RTA_tau_transport}). Dotted line represents the isotropic limit.}
 \label{fig:el_cond_mD}
\end{figure}

\begin{figure}
 \centering
 \includegraphics[scale=0.3, keepaspectratio=true]{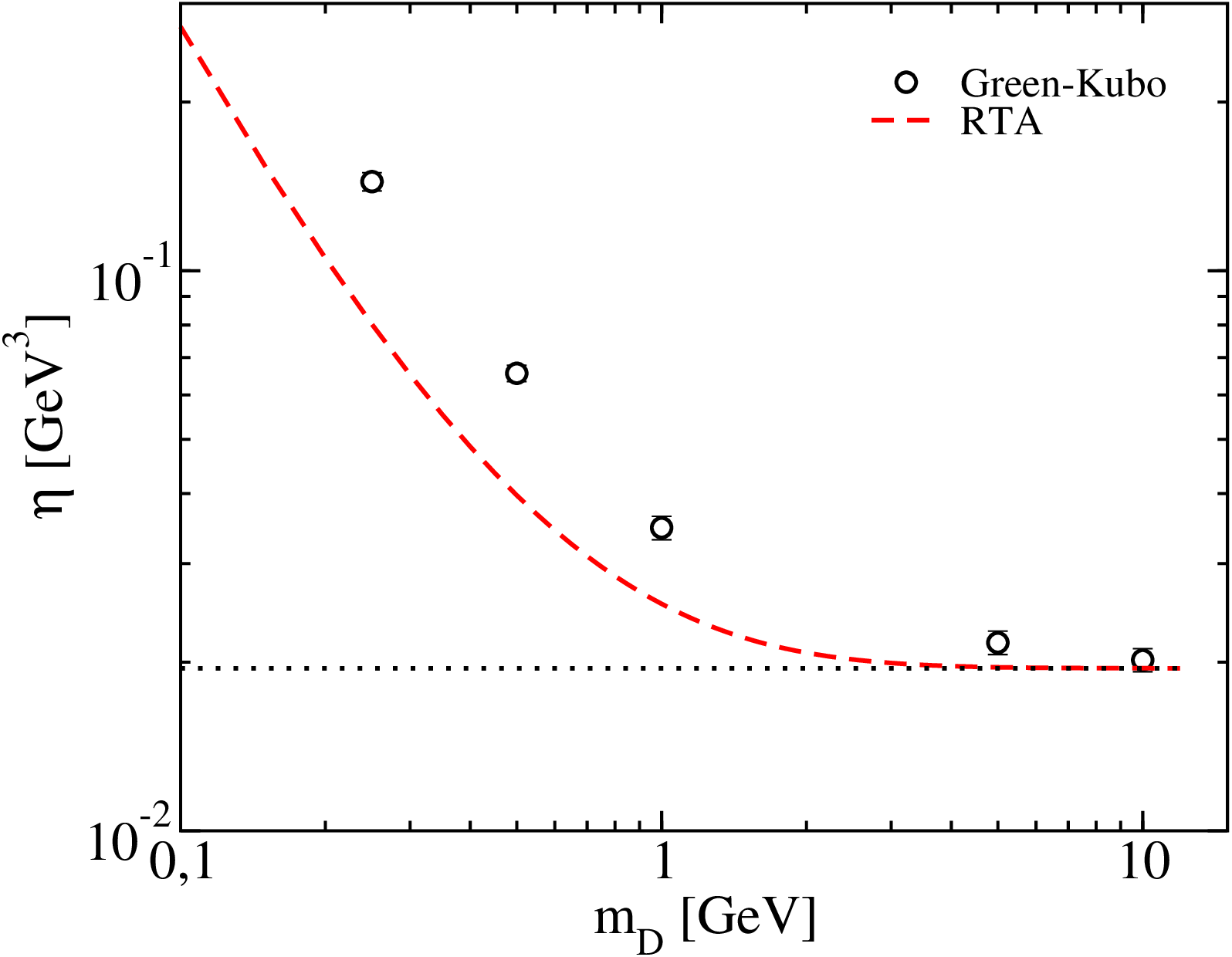}
 \caption{Shear viscosity as a function of Debye mass $m_D$. In this simulation we consider $T=0.4\,GeV$,  $\sigma_{tot}=10\,mb$ and  $m=0.4\,GeV$  for all particles. Open circles are obtained using Green-Kubo correlator.
  Dashed line represents RTA while dotted line is the isotropic limit.}
 \label{fig:eta_mD}
\end{figure}

The transport relaxation time in this case can be written as:
\begin{equation}\label{RTA_tau_transport}
\tau_{tr,i}^{-1}=\sum_j \langle \rho_j \sigma^{ij_{tr}} v_{rel}^{ij}\rangle=
\sigma_{tot} \langle v_{rel} h(a) \rangle \left( \rho_q + \rho_{\bar{q}}+ \rho_g \right)
\end{equation}
where for massive particles $\rho=\frac{\gamma}{2\pi^2} T^3 \left( \frac{m}{T}\right)^2 K_2\left( \frac{m}{T} \right)$, being $K_2$ the modified Bessel function.
We performed simulations of a system of quarks, antiquarks and gluons with $m=0.4\,GeV$, that is the same value obtained in quasi-particle model in the range of temperature $T=0.2\div 0.4\,GeV$. However at this point it represents only a systematic study: in the next section we will present quasi-particle model results with temperature dependent masses for quarks and gluons.
In Fig. \ref{fig:el_cond_mD} we show $\sigma_{el}/T$ as a function of $m_D$ for a system of quarks and gluons  interacting via the same cross-section described Eq. (\ref{sigma_anis}) with $\sigma_{tot}=10\,mb$ and $T=0.4\,GeV$: open circles represent Green-Kubo results while blue squares are obtained using the E-field method ($eE=0.05\,GeV/fm$);  red dashed line is RTA using Eq. (\ref{RTA_tau_transport}) for the relaxation time; dotted line represent the isotropic limit, i.e. $\sigma_{tr}=\frac{2}{3}\sigma_{tot}$.
We can see that RTA underestimates numerical results of $\sigma_{el}$ by a factor of about $40\%$  already at  $m_D=1\,GeV$ while in the isotropic limit $m_D>5\,GeV$ it slightly overestimates numerical estimations as we know from the simple case of massless particles of the previous section.

\begin{figure}
 \centering
 \includegraphics[scale=0.3, keepaspectratio=true]{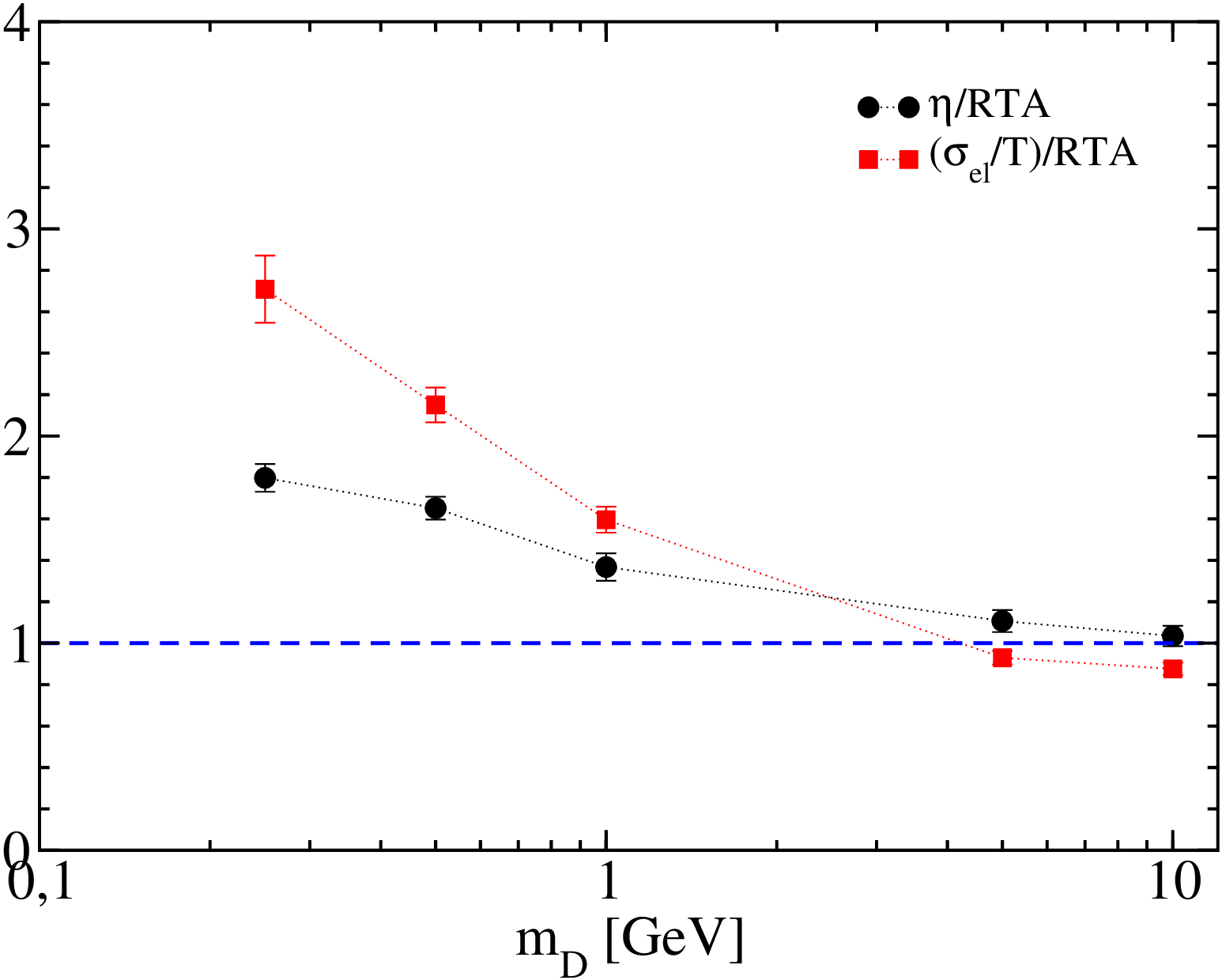}
 \caption{Green-Kubo results for shear viscosity $\eta$  and electric conductivity (from results of Fig. \ref{fig:el_cond_mD} and Fig.  \ref{fig:eta_mD}) over RTA formulas as a function of Debye mass $m_D$. RTA is a quite good approximation for isotropic scatterings while for very anisotropic cross-section $m_D<1\,GeV$ we can see that RTA underestimates in a more prominently way $\sigma_{el}/T$ than $\eta$.}
 \label{fig:eta_el_su_RTA}
\end{figure}

The behaviour of $\sigma_{el}/T$ as a function of $m_D$ is similar to the one already observed for shear viscosity in the case of massless gluons in \cite{Plumari_visco}. We show for $\eta$  the corresponding plot in Fig. \ref{fig:eta_mD} where are shown results for the same system studied for electric conductivity.
We compute shear viscosity $\eta$ using Green-Kubo relation $\eta=V/T \langle \pi^{xy}(0)^2 \rangle \tau$, as already done for single component system \cite{Plumari_visco,Wesp:2011yy, Fuini:2010xz}.
In Fig. \ref{fig:eta_mD} open circles are Green-Kubo results, dashed line is RTA and dotted line represents the isotropic limit.
As found in \cite{Plumari_visco} RTA is a good approximation for $\eta$ only for isotropic scatterings while for $m_D<1\,GeV$ it underestimates Green-Kubo results by a factor of $25\%$.

If we look in details RTA estimations for both transport coefficients we find that $\sigma_{el}$ is underestimated by RTA more considerably than $\eta$ as the cross-section becomes more forward peaked.
In Fig. \ref{fig:eta_el_su_RTA} we plot the ratio of numerical results of both transport coefficients over the analytical predictions of RTA: black circles are Green-Kubo results for $\eta$, red squares are Green-Kubo results for $\sigma_{el}/T$.
For $m_D<1\,GeV$ we can see that RTA is a better approximation for $\eta$ than for $\sigma_{el}$. 
From Fig. \ref{fig:eta_el_su_RTA} we can say that relaxation time of $\sigma_{el}$ is different from relaxation time of $\eta$ in the sense that transport relaxation time $\tau_{tr}$ in Eq. (\ref{RTA_tau_transport}) works better for $\eta$ than for $\sigma_{el}$. 
It would be interesting to study the influence of $2\to3$ inelastic scatterings to relaxation time of electric conductivity as already done for $\eta$ in \cite{Wesp:2011yy}. This study is currently pursued within the BAMPS transport approach. The effect of considering also $2\to 3$ scatterings is to decrease $\eta$ of about a factor $3$. 

\section{\label{Sec:el_cond_quasi_particle}Electric conductivity of QGP}
In this section we investigate the more realistic case of quarks, antiquarks and gluons interacting via different anisotropic and energy dependent cross-section according to the pQCD-like scheme  with a screening mass $m_D$ as arising from HTL approach: $m_D\sim g(T)T$.
The total cross-section used has the following form:
\begin{equation}
\sigma_{tot}^{ij}=\beta^{ij}\sigma(s)=\beta^{ij}\frac{\pi \alpha^2_s}{m_D^2}\frac{s}{s+m_D^2}
\end{equation}
where $\alpha_s=g^2/4\pi$ and the coefficient $\beta^{ij}$ depends on the species of interacting particles: $\beta^{qq}=16/9$, $\beta^{qq'}=8/9$, $\beta^{qg}=2$, $\beta^{gg}=9$.

\begin{figure}
 \centering
 \includegraphics[scale=0.28, keepaspectratio=true]{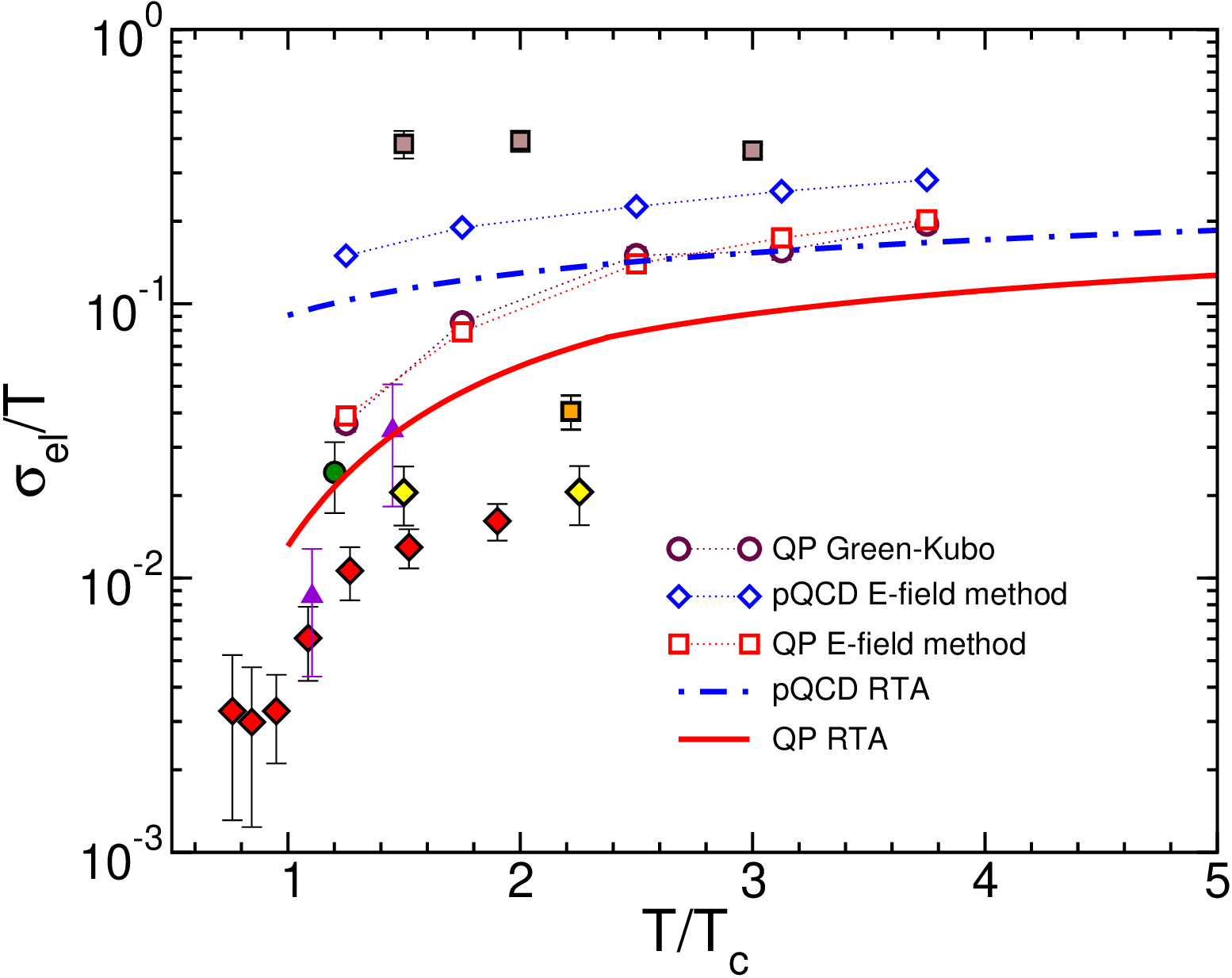}
 \caption{Electric conductivity $\sigma_{el}/T$ as a function of $T/T_c$. Dark open circles are Green-Kubo results for QP model, red open squares are QP model results computed with the E-field method, violet open diamonds are pQCD results calculated with the E-field method; red line  and violet line are RTA predictions respectively for the QP model and pQCD case. Symbols are Lattice data: grey squares \cite{Gupta}, 
violet triangles \cite{Ding}, green circle \cite{Brandt}, yellow diamonds \cite{Aarts}, red diamonds\cite{Amato} and orange square\cite{Buividovich}. }
 \label{fig:el_cond_qp}
\end{figure}

Furtheremore, in order to take into account the thermodynamics from lQCD calculation, we employ the quasi-particle (QP) model \cite{Plumari_qpmodel} similarly to \cite{Levai:1997yx,Peshier:2002ww,Bluhm:2010qf,Bluhm:2004xn}. We recall that the aim of a quasi-particle model is to describe a strongly interacting systems in terms of quasi-particles weakly interacting whose masses are generated by the non-perturbative effects.
The QP model, as a phenomenological way to describe microscopically the QGP, has become a quite solid approach  for $T>2-3\,T_c$ specifically using NNLO HTLpt \cite{Andersen:2010wu,Blaizot}.
We notice that a dynamical quasi-particle model (DQPM), which includes also spectral functions, has been developed in \cite{Cassing:2009vt} where a very similar behaviour of $g(T)$ has been deduced. The main difference between QP and DQPM comes from the fact that DQPM considers isotropic scatterings that, according to our results of previous section, decrease electric conductivity of about $20-30\%$ \cite{Cassing_el} in the range of anisotropy of interest as we will discuss below.
In Ref. \cite{Plumari_qpmodel}, performing a fit to the lattice energy density,  we have obtained the following parametrization for the running coupling:
\begin{equation}
g^2(T)=\frac{48 \pi^2}{\left( 11 N_c - 2 N_f\right) \ln\left[ \lambda \left( \frac{T}{T_c}-\frac{T_s}{T_c} \right) \right]^2}
\end{equation}
with $\lambda = 2.6$ and  $T_s/T_c=0.57$. We notice that such a fit reproduce the exact result with a very good precision only for $T> 1.1\, T_c$. Quarks and gluons masses are given by $m_g^2=3/4 g^2T^2$ and $m_q^2=1/3g^2T^2$.

We also study the behaviour of electric conductivity using the pQCD running coupling $g_{pQCD}=\frac{8\pi}{9}\ln^{-1}\left( \frac{2\pi T}{\Lambda_{QCD}} \right)$ considering massless particles: even if this case is not able to describe the phase transition, it is interesting to study the $\sigma_{el}$ dependence on a different running coupling and also to consider an asymptotic limit valid for $T\gg T_c$.

In Fig. \ref{fig:el_cond_qp}, we show electric conductivity $\sigma_{el}/T$ as a function of $T/T_c$. Open circles are computed using Green-Kubo correlator, red open squares with the E-field method (applying an $eE=0.02\div 0.05\, GeV/fm$ to guarantee the saturation of electric current) for the quasi-particle model, blue  open diamonds represent result for the massless pQCD case that we have computed only with the E-field method. Red line is RTA for QP model, blue dot-dashed  line RTA for the massless pQCD case. Symbols denotes Lattice data: grey squares \cite{Gupta}, 
violet triangles \cite{Ding}, green circle \cite{Brandt}, yellow diamonds \cite{Aarts}, red diamonds\cite{Amato} and orange square\cite{Buividovich}.
Also in this case Green-Kubo results are in good agreement with E-field method in the range of temperature explored.
Numerical results predicted by the QP model are about a factor of $4$ greater then recent Lattice QCD calculations \cite{Amato}.
As we discussed in \cite{Puglisi}, also $\eta/s$ predicted by the QP model is about a factor $4-5$ greater than the minimum value $1/4\pi$ near $T_c$. This means that rescaling $\eta/s$, in order to reproduce  the minimum value, one obtains an electric conductivity $\sigma_{el}$ very close to recent Lattice data \cite{Amato}.

\begin{figure}
 \centering
 \includegraphics[scale=0.3, keepaspectratio=true]{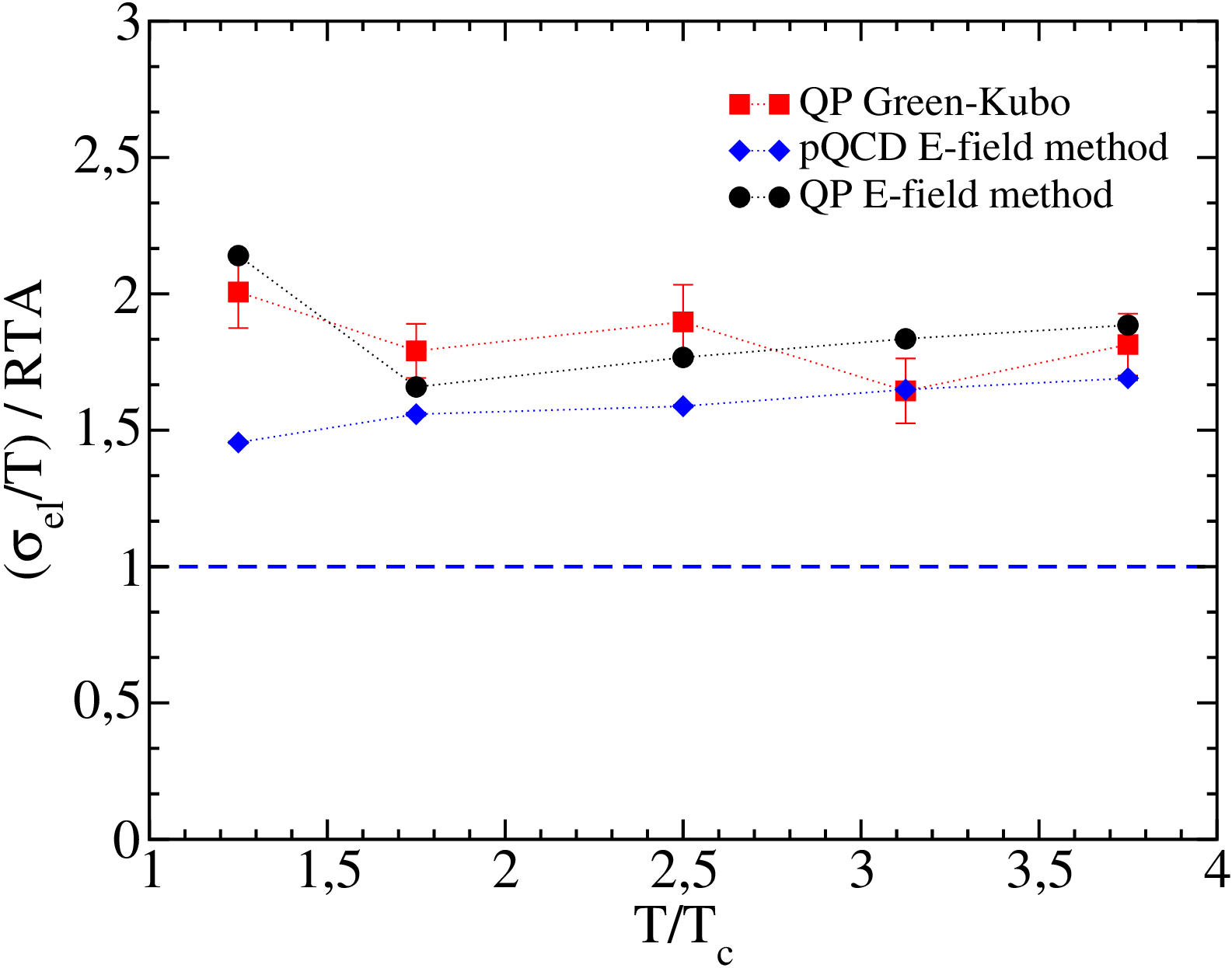}
 \caption{Ratio between numerical results of $\sigma_{el}/T$ and RTA predictions as a function of $T/T_c$: red squares are obtained using Green-Kubo results for QP model, dark circles with E-field method for QP model, blue diamonds with E-field method for the pQCD case.}
 \label{fig:QP_el_su_RTA}
\end{figure}

We note again that RTA underestimates $\sigma_{el}/T$ for both QP model and pQCD case, as we could expect qualitatively  from the previous section of  anisotropic scattering considering that in the QP model $m_D\simeq 0.8-1.2\,GeV$  and in the pQCD case $m_D\simeq 0.4-1\,GeV$ for temperature $T=0.2-0.6\,GeV$.
We can see in details in Fig. \ref{fig:QP_el_su_RTA} the ratio between numerical results of electric conductivity $\sigma_{el}/T$ and RTA predictions as a function of $T/T_c$. Red squares are calculated taking the ratio between Green-Kubo results and RTA estimation for the QP model, black circles with the E-field method for QP model while blue diamonds with the E-field method for the pQCD case. We notice that QP results are underestimated by an average factor of $1.8$ and pQCD results by $1.6$ in the range of temperature explored. 

\section{Conclusion}
Transport coefficients, like shear and bulk viscosity, heat and electric conductivity, characterize the response of a system to different kind of  perturbations and regulate the dynamics of the system toward the equilibrium state through dissipation.
Being the QGP created in Heavy Ions Collisions a system far from equilibrium, the study and evaluation of transport coefficients for strongly interacting matter is mandatory.
In this work we have computed electric conductivity solving numerically the Relativistic Boltzmann Transport equation with a parton cascade code already developed using two methods: the Electric field method, suggested by the definition itself of $\sigma_{el}$ ($J=\sigma_{el}E$), and Green-Kubo correlator.
We have studied in a systematic way the microscopic details dependence of $\sigma_{el}$ on the particular scattering (isotropic and anisotropic cross-section).
We have found that RTA formula of $\sigma_{el}$ is a  quite good approximation for massless and massive particles interacting with isotropic cross section. We observe that it overestimates numerical results by about $30\%$ for temperature $T>0.5\,GeV$.
For massive quarks, antiquarks and gluons interacting via anisotropic cross-section we have calculated $\sigma_{el}$ and shear viscosity $\eta$ showing the differences between the two transport coefficients: RTA underestimates  both $\sigma_{el}$ and $\eta$. However we show that RTA prediction is worse for $\sigma_{el}$ than for $\eta$ meaning that relaxation time $\tau_{\sigma_{el}}$ is not exactly equal to $\tau_{\eta}$.

We find two main results. The first is that Green-Kubo and the E-field method to estimate the electric conductivity agree quite well in all the range of temperature explored that are those relevant for the hot QGP physics. The second result is that the RTA prediction of $\sigma_{el}$ based on relaxation time given by the transport cross-section always underestimates the conductivity. Such a discrepancy increases for more forward peaked cross-section (small $m_D$). We find for $m_D\sim T$ that such a discrepancy can be about a factor of $2$. Instead for more realistic description with running strong coupling we find a discrepancy of about $60\%$ for both the pQCD and QP description.

Our results is of  wide impact and shows that when a microscopic description is employed to estimate the $\sigma_{el}$ of the QGP, or of hadronic matter, if the $\tau_{tr}$ is used then one may expect to significantly underestimate $\sigma_{el}$.
Our results is quite general but it has been found considering only a system that scatters with $2\leftrightarrow 2$ collisions. It is very interesting to have similar information also when $2\leftrightarrow 3$ collisions are included as in BAMPS \cite{Wesp:2011yy} and also to compute bulk viscosity \cite{Marty} and heat conductivity \cite{Greif} that are the other two transport coefficients of interest in an expanding quark-gluon plasma.

\section{Acknowledgments}
The authors thank C. Greiner and M. Greif for useful comments and discussions.
A. P. thanks C. Greiner and M. Greif for the courteous hospitality at Johann Wolfgang Goethe-Universit\"at  in Frankfurt  that stimulated numerical comparisons between our transport codes.
V. G. acknowledges the support by the ERC-StG under the QGPDyn grant.


\end{document}